\documentclass[conference]{IEEEtran}
\IEEEoverridecommandlockouts
\usepackage{enumitem}
\usepackage{amssymb}
\usepackage{dutchcal}

\usepackage{amsthm} 
\newtheorem{theorem}{Theorem}

\usepackage{graphicx}
\usepackage{subfigure}
\UseRawInputEncoding

\usepackage{cite}
\usepackage{amsmath,amssymb,amsfonts}
\usepackage{algorithmic}
\usepackage{graphicx}
\usepackage{textcomp}
\usepackage{xcolor}
\def\BibTeX{{\rm B\kern-.05em{\sc i\kern-.025em b}\kern-.08em
    T\kern-.1667em\lower.7ex\hbox{E}\kern-.125emX}}
\begin{document}

\title{On 1-Persistent Retransmission in Grant-free Access for URLLC Service

\author{
	Dannan~Hong,
	
	Tong~Ye,~\IEEEmembership{Member,~IEEE}
	
	\thanks{Dannan Hong, Tong Ye are with the State Key Laboratory of Advanced Optical Communication Systems and Networks, Shanghai Jiao Tong University, Shanghai 200240, China (e-mail: hdn0514@sjtu.edu.cn; yetong@sjtu.edu.cn).}}%

}



\maketitle

\begin{abstract}
Ultra-reliable and low-latency communication (URLLC) is one of three major application scenarios of the 5G new radio, which has strict latency and reliability requirements. Contention-based grant-free (GF) access protocols, such as Reactive, K-Repetition, and Proactive, have been proposed for uplink URLLC service. In the GF access, user equipment (UE) resends packet immediately after an unsuccessful transmission such that the latency requirement can be satisfied. Taking Reactive as an example, this paper studies the impact of 1-persistent retransmission (1-pR) on the distribution of user-plane delay. We define the number of UEs that try to send packets in each mini-slot as attempt rate. We show that the 1-pR makes the attempt rate seen by the packet in retransmission larger than that seen by the packet in the first transmission. As a result, the successful probability of retransmission is lower than that of the first transmission. Based on this observation, we derive the distribution of user-plane delay, which also takes into account the delay incurred by queueing process. We demonstrate that whether to include the effect of 1-pR and queueing process in the analysis would have a significant impact on the prediction accuracy of delay distribution.
\end{abstract}

\begin{IEEEkeywords}
Ultra-reliable and low-latency communication, grant-free access, 1-persistent retransmission, user-plane delay.
\end{IEEEkeywords}

\section{Introduction}\label{sec_Intro}
Ultra-reliable and low-latency communication (URLLC) is an important application in future wireless networks. One of the design purposes is to provide communication service for a class of sporadic event-driven traffic \cite{uRLLC_sporadic}, which has very strict requirements on latency and reliability. According to the 3GPP standard \cite{3gpp_standard}, the user-plane delay, which is defined as the time from when a packet arrives at a user equipment (UE) to when it is successfully received by the base station (BS), should be less than 1ms with probability 0.99999.

To meet such requirements, several grant-free (GF) access protocols, such as Reactive, K-Repetition, and Proactive \cite{Grant_free_class} have been proposed for URLLC service. Different from the traditional grant-based (GB) access protocol for enhanced mobile broadband (eMBB), the GF protocols do not require the UE to request resource blocks (RBs) from the BS before data transmission. The wireless access network is a slotted system, where the duration of a mini-slot is called transmission time interval (TTI). A typical TTI is 1/7ms \cite{0.143_figure}. If a packet arrives in a TTI, the UE randomly selects an RB and transmits the packet to the BS at the beginning of the next TTI, without the need for a time-consuming handshake process similar to that in GB protocols. Compared to the GB access, the GF scheme can eliminate a $\sim$10-ms delay overhead \cite{liuyan}. Thus, the GF access mode has been included in the 3GPP standard \cite{3gpp_standard} to cope with URLLC traffic in the uplink direction.

However, the GF access will inevitably lead to data collisions, due to the lack of coordination. Because of very strict latency requirement of URLLC traffic, a very-small collision probability may have a significant impact on the reliability of transmission. After a UE starts to send a packet, it will receive an acknowledgment (ACK) or negative acknowledgment (NACK) from the BS in the 3-th TTI, depending on whether the transmission is successful or not. To ensure the user-plane delay is lower than 1ms as much as possible in the case where the transmission fails, the UE has to retransmit the packet in the next TTI with probability 1, called 1-persistent retransmission (1-pR) in this paper. A similar concept is the 1-persistent carrier sense multiple access (CSMA) \cite{1_CSMA}. To meet the delay requirement, the network should assign a proper number of RBs in each TTI to the URLLC traffic according to the population of UEs in the cell, such that a packet can be received by the BS successfully via one or two transmissions with probability 0.99999. Therefore, understanding the transmission mechanism of GF access would be significant to proper wireless spectrum allocation.

To serve this purpose, several analytical models \cite{liuyan,similat_liuyan_reactive,similat_liuyan_K_rep,similat_liuyan_MIMO,Markov_chain_proactive} have been developed to analyze the delay performance of GF access. Ref. \cite{liuyan} proposes an iterative approach to derive the outage probability, which is defined as the probability that the user-plane delay is higher than the delay requirement. Based on the method reported in \cite{liuyan}, Refs. \cite{similat_liuyan_reactive} and \cite{similat_liuyan_K_rep} solve the outage probability, taking the finite block length coding theory into consideration. In the context of millimeter wave (mmWave) and massive multiple input multiple output (MIMO), Ref. \cite{similat_liuyan_MIMO} obtains a closed-form approximation of latent access failure probability, using stochastic geometric spatiotemporal tools. Ref. \cite{Markov_chain_proactive} models the transmission process of each UE as a discrete-time Markov process, deriving packet loss probability to characterize reliability. Aside from ignoring the queueing process of packets, most of these works assume that a UE can randomly move to the coverage of an arbitrary BS in different TTIs, i.e., the location of UEs follows a Poisson point process (PPP) \cite{liuyan,similat_liuyan_reactive,similat_liuyan_K_rep,similat_liuyan_MIMO}. The data collision of UEs in a TTI is then considered independent of that in other TTIs. However, this assumption is too ideal to be consistent with practical situations.

As Ref. \cite{move_infnite} points out, the UE positions in different TTIs are correlated. The moving speed of the UE is limited. For example, in urban areas, the velocity of a car is 30-60km/h, while that of a pedestrian is 3-5km/h. Moreover, the UEs in some scenarios, e.g., the sensor installed in factories \cite{sensor,industrial_Monitoring_Sensors}, may even keep unmoved. This means a UE may stay in the coverage of the same BS during a long period. Under this condition, the 1-pR in GF access will introduce a dependency among the data collisions in different TTIs and such a dependency will remarkably affect the delay distribution of packets, as section \ref{sec_sysModel} will show.

\subsection{Our Work}
In this paper, we study the effect of 1-pR on the collision process of GF access, under the condition that the group of UEs in the coverage of a BS is relatively fixed. Though there are three strategies for GF access, we model the Reactive strategy \cite{Grant_free_class} in this paper. We define the number of UEs that try to send packets as the attempt rate. We show that the 1-pR makes the attempt rate seen by the packet in retransmission larger than that seen by the packet in the first transmission. As a result, the successful probability of retransmission is lower than that of the first transmission. Taking this observation into consideration, we derive the distribution of the user-plane delay of URLLC packets, which includes the delay incurred by queueing process. We demonstrate that whether to include the effect of 1-pR and queueing process in the analysis would have a significant impact on the prediction accuracy of delay distribution. We further show that our analytical result can be used to predict the maximum number of UEs that can be supported, if other parameters, such as the number of available RBs and the traffic rate are known. 

In summary, the contributions of this paper are as follows:
\begin{enumerate}
    \item We study the impact of the 1-pR of GF access on user-plane latency in the case where the group of UEs in the coverage of a BS is relatively fixed.
    \item Different from previous works \cite{liuyan,similat_liuyan_reactive,similat_liuyan_K_rep,similat_liuyan_MIMO,Markov_chain_proactive}, we take into account the queueing process of packets in our model and show that it has a significant influence on delay performance.
    \item Our result can be used to predict the number of UEs that can be supported when the number of RBs is known.
\end{enumerate}

The rest of this paper is organized as follows. Section \ref{sec_sysModel} presents the system model and our assumptions. Section \ref{sec_1pR} analyzes the impact of 1-pR on the attempt rates seen by the packets in the first transmission and retransmission. Taking into account the queueing process of packets, we derive the distribution of user-plane delay in section \ref{sec_Delay}, based on which section \ref{sec_discussion} demonstrates the impact of 1-pR and queueing process on delay performance and system design. Section \ref{sec_conclusion} concludes this paper.

\section{System Model and Assumptions}\label{sec_sysModel}
In this paper, we study the uplink transmission of URLLC traffic via GF access in the coverage of a BS. In the network, $N$ UEs compete for $B$ RBs in each TTI. The arrival rate of URLLC packets to the access network is $\hat{\lambda}$ packets/s, and to each UE is $\lambda=\hat{\lambda} / N$ packets/s.

This paper considers a fading channel, which has an additive noise with an average power of $\sigma^{2}$ and experiences frequency-flat Rayleigh fading. All UEs adopt full path-loss power control to compensate for the “near/far” effect \cite{near_far}, such that the packets from different UEs have the same mean received power, denoted by $\bar{P}$, at the BS. Consider a packet $j$, which selects the same RB with a set of other packets at a TTI, denoted by $Q_{j}$. The signal to noise plus interference ratio (SINR) of packet $j$ at the BS is given by
\begin{equation}
SINR=\frac{P_{j}}{\sigma^{2}+\sum_{i \in Q_{j}} P_{i}},\label{eq_SINR}
\end{equation}
where $P_{j}=\bar{P}\left|h_{j}\right|^{2}$ is instantaneous received power and $\left|h_{j}\right|^{2} \sim \operatorname{Exp}(1)$ is channel power gain.

Sometimes, we also consider the ideal channel, which has no additive noise. In this case, the transmission fails only due to packet collision. The discussion of the situation of the ideal wireless channel helps to better demonstrate the effect of 1-pR on the transmission process of packets.

We take into consideration the capture effect \cite{capture_model} of the radio receiver at the BS in our model. With the capture effect, the radio receiver can independently decode each UE's packet by treating others as background noise, when several packets compete for the same RB. The receiver can decode a packet successfully, as long as the received SINR is above a certain threshold, denoted by $\mu$.

\begin{figure}[b]
\centerline{\includegraphics[width=0.85\linewidth]{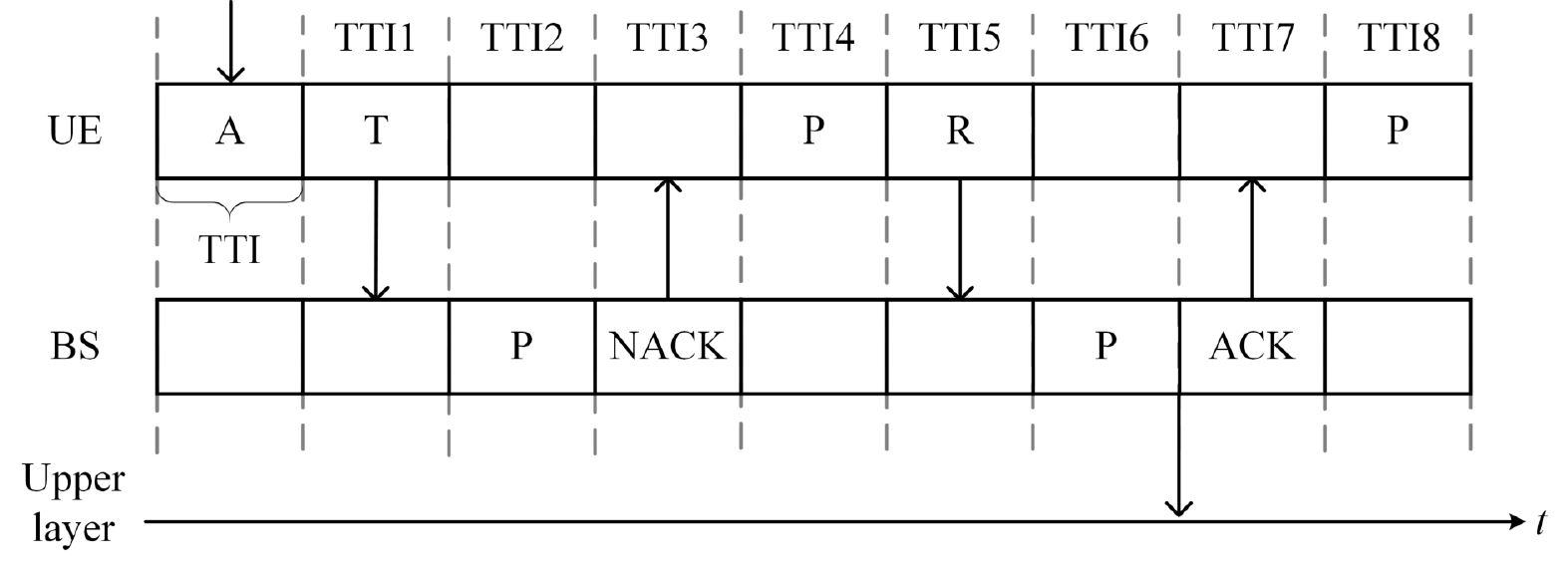}}
\caption{Reactive strategy, where A, T, P, and R stand for (A)lignment, (T)ransmission, (P)rocessing, and (R)etransmission, respectively \cite{0.143_figure}.}
\label{fig_Reactive}
\end{figure}

The UEs compete for the RBs using the Reactive strategy. As Fig. \ref{fig_Reactive} plots, after a packet arrives at the UE, it first experiences an alignment delay. The UE randomly selects an RB to transmit the packet at the beginning of the next TTI, called TTI 1. At the end of TTI 1, the BS receives the packet. After that, the BS takes one TTI to process the packet. If the BS decodes the packet successfully, the BS submits the packet to the upper protocol layer and feeds back an ACK to the UE. The BS may fail to decode the packet due to insufficient SINR induced by channel fading or data collision. In this case, the BS sends NACK to the UE, which costs one TTI also. The UE spends one TTI processing the NACK and retransmits the packet at the start of TTI 5, which is called 1-pR in this paper.

In addition to the above description, we adopt the following assumptions in our analysis:

\begin{enumerate}[label=A\arabic*]
\item 	The number of UEs within the coverage of the BS $N$ is sufficiently large.
\item   The group of UEs keeps unchanged under the coverage of a BS during a large number of TTIs.
\item 	The total arrival rate of the URLLC traffic input to the network $\hat{\lambda}$ is not large, such that a high successful probability of packet transmission can be ensured.
\item  The arrival process of packets to each UE is a Poisson process with rate $\lambda$.
\item  The number of RBs assigned for URLLC service $B$ is sufficiently large.
\item The retransmission continues until a packet is successfully received by the BS.
   
\end{enumerate}

\section{Effect of 1-Persistent Retransmission} \label{sec_1pR}
Given that the UE group under the coverage of a BS keeps unchanged during a long period, the 1-pR has an effect on the competition process of URLLC. Define the number of UEs that attempt to transmit packets in each TTI as attempt rate. Consider a UE, denoted by $\mathcal{A}$. When it transmits the head-of-line (HOL) packet for the first time, whether or not the remaining $N-1$ UEs send packets is independent of $\mathcal{A}$. One of possibilities that the transmission fails is that the packet of $\mathcal{A}$ competes with those of other nodes for the same RB. When $\mathcal{A}$ retransmits the packet, the UEs that fail in the competition with $\mathcal{A}$ will reattempt definitely. For example, the packets of UE 1 and UE 2 collide in the same RB and their transmissions fail in TTI 1, and both of them will be resent in TTI 5. Intuitively, this will make the attempt rate seen by the packet of $\mathcal{A}$ in retransmission larger than that seen by the packet in the first transmission.

\begin{figure}[t]
\centerline{\includegraphics[width=0.85\linewidth]{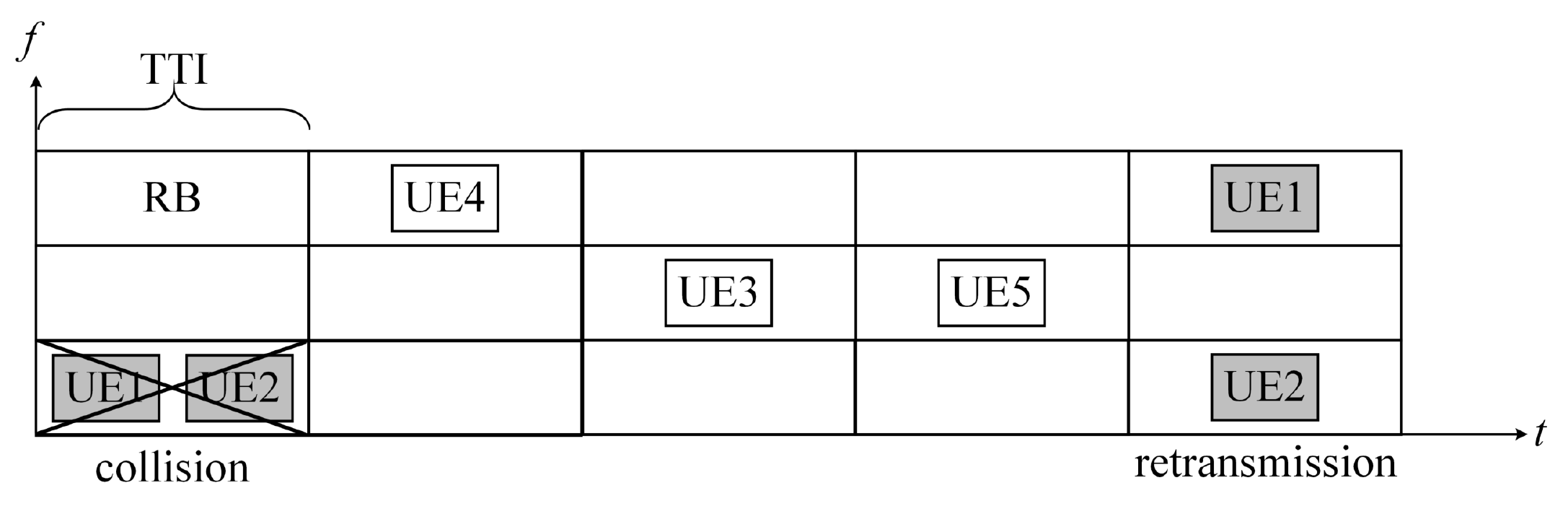}}
\caption{The 1-pR in the Reactive strategy.}
\label{fig_1pR}
\end{figure}

Let $G_{F}$ and $G_{R_{i}}$ be the attempt rates seen by the HOL packet in the first transmission and the $i$-th retransmission, respectively, where $i=1,2, \cdots$. In this section, we show that $G_{F}<G_{R_{1}} \approx G_{R_{2}} \approx \cdots$ in most cases, due to the 1-pR.

\subsection{Transmission of HOL Packet}
To derive the attempt rates, we need to study the competition process that an HOL packet will experience before it is successfully received by the BS. After a packet becomes an HOL packet, it experiences the first transmission lasting for 1 TTI, as Fig. \ref{fig_1pR} shows. After that, it waits 3 TTIs in the buffer. If the transmission fails, it will undergo a retransmission in the 5-th TTIs after the first transmission. This process continues until this packet is received by the BS successfully.

\begin{figure}[t]
\centerline{\includegraphics[width=\linewidth]{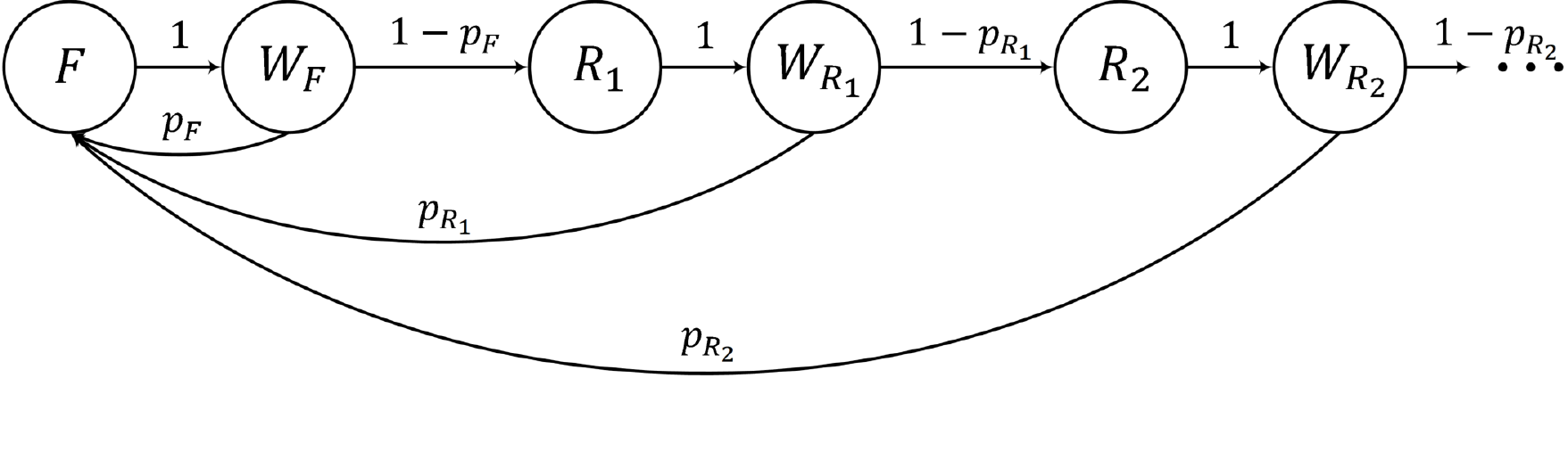}}
\caption{Markov chain of the competition process of an HOL packet.}
\label{fig_MC}
\end{figure}

As Fig. \ref{fig_MC} plots, the transmission process of an HOL packet can be described by a Markov chain, where $F$ and $R_{i}$ are the state of the first transmission and the state of the $i$-th retransmission, $W_{F}$ and $W_{R_{i}}$ are the waiting states after the first transmission and the $i$-th retransmission, and $p_{F}$ and $p_{R_{i}}$ are the successful probabilities of the first transmission and the $i$-th retransmission, respectively. Clearly, the duration times of $F$ and $R_{i}$ are 1 TTI, and those of $W_{F}$ and $W_{R_{i}}$ are 3 TTIs. Let $\pi_{F}$, $\pi_{W_{F}}$, $\pi_{R_{i}}$, and $\pi_{W_{R_{i}}}$ be the limiting probabilities of states $F$, $W_{F}$, $R_{i}$, and $W_{R_{i}}$. According to Fig. \ref{fig_MC}, we have the following equilibrium equations:
\[\pi_{F}=p_{F} \pi_{W_{F}}+\sum_{j=1}^{\infty} p_{R_{j}} \pi_{W_{R_{j}}},\]
\[\pi_{W_{F}}=\pi_{F},\]
\[\pi_{W_{R_{i}}}=\pi_{R_{i}},\]
\[\pi_{R_{1}}=\left(1-p_{F}\right) \pi_{W_{F}},\]
\[\pi_{R_{i+1}}=\left(1-p_{R_{i}}\right) \pi_{W_{R_{i}}}.\]
Solving them together with the normalization condition
\[\pi_{F}+\pi_{W_{F}}+\sum_{j=1}^{\infty}\left(\pi_{R_{j}}+\pi_{W_{R_{j}}}\right)=1,\]
we have the limiting probabilities as follows
\begin{equation}
\left\{\begin{array}{c}
\pi_{F}=\pi_{W_{F}}=\frac{1}{2\left[1+\left(1-p_{F}\right)+\left(1-p_{F}\right) \sum_{i=2}^{\infty} \prod_{j=1}^{i-1}\left(1-p_{R_{j}}\right)\right]},\\
\pi_{R_{i}}=\pi_{W_{R_{i}}}=\frac{\left(1-p_{F}\right) \prod_{j=1}^{i-1}\left(1-p_{R_{j}}\right)}{2\left[1+\left(1-p_{F}\right)+\left(1-p_{F}\right) \sum_{i=2}^{\infty} \prod_{j=1}^{i-1}\left(1-p_{R_{j}}\right)\right]}.\\
\end{array}\right.\label{eq_LimitPro}
\end{equation}
Let $\tilde{f}$ and $\tilde{r}_i$ be the time-average probabilities that the UE stays at states $F$ and $R_{i}$ during the time interval when the UE is busy with HOL-packet transmission. From (\ref{eq_LimitPro}), we can easily obtain $\tilde{f}$ and $\tilde{r}_i$ as follows
\begin{equation}
\tilde{f}=\frac{\pi_{F}}{\pi_{F}+3 \pi_{W_{F}}+\sum_{j=1}^{\infty} \pi_{R_{j}}+3 \pi_{W_{R_{j}}}}=\frac{1}{2} \pi_{F},\label{eq_f}
\end{equation}
and
\begin{equation}
\tilde{r}_{i}=\frac{\pi_{\mathrm{R}_{i}}}{\pi_{F}+3 \pi_{W_{F}}+\sum_{j=1}^{\infty} \pi_{R_{j}}+3 \pi_{W_{R_{j}}}}=\frac{1}{2} \pi_{R_{i}}.\nonumber
\end{equation}
It follows that the probability that a UE is non-empty, denoted by $\alpha$, is given by \cite{non_empty_probability}   
\begin{equation}
\alpha=\lambda \tilde{f}^{-1}.\label{eq_alpha}
\end{equation}

\subsection{Attempt Rates}
When UE $\mathcal{A}$ transmits the HOL packet for the first time, the average number of non-empty UEs seen by UE $\mathcal{A}$ is $(N-1) \alpha$. Also, the probability that a UE is busy in transmission is $\tilde{f}+\sum_{i=1}^{\infty} \tilde{r}_{i}$. The attempt rate seen by the packet in the first transmission is thus given by
\begin{equation}
G_{F}=(N-1) \alpha\left(\tilde{f}+\sum_{i=1}^{\infty} \tilde{r}_{i}\right)=\frac{\alpha}{4}(N-1).\label{eq_GF}
\end{equation}

Intuitively, the attempt rate seen by the packet in the $i$-th retransmission $G_{R_{i}}$ depends on the attempt rate seen by the packet in the $(i-1)$-th retransmission $G_{R_{i-1}}$. In the following, we first derive $G_{R_{1}}$ from $G_{F}$, and then calculate $G_{R_{i}}$ from $G_{R_{i-1}}$, where $i=2,3,\cdots$.

$G_{R_{1}}$ consists of two parts. Let $g_{1}$ be the average number of UEs that do not compete for the same RB with UE $\mathcal{A}$ in $\mathcal{A}$'s first transmission and send packets in $\mathcal{A}$'s first retransmission, and $g_{2}$ be the average number of UEs that fail in the competition with UE $\mathcal{A}$ in $\mathcal{A}$'s first transmission. Clearly, $G_{R_{1}}=g_{1}+g_{2}$.

Before calculating $g_{1}$ and $g_{2}$, we need to know the condition probability $\operatorname{Pr}\left\{S_{k} \mid H\right\}$, where $S_{k}$ is the event that $k$ UEs compete for the same RB with $\mathcal{A}$, and $H$ is the event that $\mathcal{A}$ fails in the first transmission of HOL packet. According to the Bayes' theorem
\begin{equation}
\operatorname{Pr}\left\{S_{k} \mid H\right\}=\frac{\operatorname{Pr}\left\{H \mid S_{k}\right\} \operatorname{Pr}\left\{S_{k}\right\}}{\operatorname{Pr}\{H\}}.\label{eq_SkH}
\end{equation}
When $N$ is sufficiently large, it is easy to show that the number of UEs sending packets in a TTI obeys the Poisson distribution, that is,
\begin{equation}
\operatorname{Pr}\left\{S_{k}\right\}=\frac{\left(\frac{G_{F}}{B}\right)^{k} e^{-\frac{G_{F}}{B}}}{k !}.\label{eq_Sk}
\end{equation}
According to Ref. \cite{k_UE_success_probability}, the probability that $\mathcal{A}$ fails in the first transmission under the condition that $k$ UEs compete for the same RB with $\mathcal{A}$ is given by   
\begin{equation}
\operatorname{Pr}\left\{H \mid S_{k}\right\}=1-\frac{e^{-\frac{\mu}{\rho}}}{(\mu+1)^{k}},\label{eq_HSk}
\end{equation}
where $\rho=\bar{P} / \sigma^{2}$ is mean received signal-to-noise ratio (SNR). Also, $\operatorname{Pr}\{H\}$ can be determined via the total probability formula as follows
\begin{align}
\operatorname{Pr}\{H\}&=\sum_{k=0}^{N-1} \operatorname{Pr}\left\{H \mid S_{k}\right\} \operatorname{Pr}\left\{S_{k}\right\}\nonumber\\
&\rightarrow \sum_{k=0}^{\infty} \operatorname{Pr}\left\{H \mid S_{k}\right\} \operatorname{Pr}\left\{S_{k}\right\},\label{eq_H}
\end{align}
where we use the condition that $N$ is sufficiently large for the second step. We verify via simulation that the difference between the second and the third terms of (\ref{eq_H}) is less than $10^{-76}$, when $N\ge20$ and $B=20$. Substituting (\ref{eq_Sk}) through (\ref{eq_H}) into (\ref{eq_SkH}), we have
\begin{equation}
\operatorname{Pr}\left\{S_{k} \mid H\right\}=\frac{\frac{\left(\frac{G_{F}}{B}\right)^{k} e^{-\frac{G_{F}}{B}}}{k !}\left[1-\frac{e^{-\frac{\mu}{\rho}}}{(\mu+1)^{k}}\right]}{1-e^{-\frac{\mu} {\mu+1} \frac{G_{F}}{B}-\frac{\mu}{\rho}}}.\label{eq_SkH_1}
\end{equation}
Consider the term $G_{F} / B$ in (\ref{eq_SkH_1}). As Eq. (\ref{eq_alpha}) shows, $G_{F} \rightarrow \hat{\lambda} /(4 \tilde{f})$ when $N$ is sufficiently large. Also, the network must keep the successful probability of packet transmission very high to meet the strict delay requirement. This implies that $\tilde{f}$ in (\ref{eq_f}) is approximately equal to 1/4. It follows that $G_{F}$ is almost equal to $\hat{\lambda}$. According to assumptions A3 and A5, $G_{F} / B$ will be very small, such that $e^{-G_{F} / B} \rightarrow 1-G_{F} / B$. Our simulation verifies that $G_{F}=0.0286$ and the difference $e^{-G_{F} / B}-(1-G_{F}/B)$ is less than $10^{-6}$ when $B \geq 20$ and $N=20$. Therefore, (\ref{eq_SkH_1}) can be rewrote as follows
\begin{equation}
\operatorname{Pr}\left\{S_{k} \mid H\right\}=\frac{\frac{\left(\frac{G_{F}}{B}\right)^{k}}{k !}\left[1-\frac{e^{-\frac{\mu}{\rho}}}{(\mu+1)^{k}}\right]}{1-e^{-\frac{\mu}{\rho}}\left(1-\frac{\mu}{\mu+1} \frac{G_{F}}{B}\right)}.\label{eq_SkH_2}
\end{equation}
From (\ref{eq_SkH_2}), we can find that
\begin{enumerate}[label=\arabic*)]
\item 	When $k=0$, 
\begin{equation}
\operatorname{Pr}\left\{S_{0} \mid H\right\}=\frac{1-e^{-\frac{\mu}{\rho}}}{1-e^{-\frac{\mu}{\rho}}\left(1-\frac{\mu}{\mu+1} \frac{G_{F}}{B}\right)}.\nonumber\end{equation}
\item   When $k=1$, 
\begin{equation}
\operatorname{Pr}\left\{S_{1} \mid H\right\}=\frac{\frac{G_{F}}{B}\left(1-\frac{e^{-\frac{\mu}{\rho}}}{\mu+1}\right)}{1-e^{-\frac{\mu}{\rho}}\left(1-\frac{\mu}{\mu+1} \frac{G_{F}}{B}\right)}.\nonumber
\end{equation}
\item 	When $k\geq 2$, $\operatorname{Pr}\left\{S_{k} \mid H\right\}$ becomes the high-order infinitesimal of $G_{F} / B$, indicating that the failure of $\mathcal{A}$ in the first transmission is induced by the collision with more than one UE with negligible probability.
\end{enumerate}
In other words, when $B$ is sufficiently large, $\operatorname{Pr}\left\{S_{k} \mid H\right\} \rightarrow 0$ for $k\geq 2$.

Let $s$ be the probability that a UE competes for the same RB with UE $\mathcal{A}$ in $\mathcal{A}$'s first transmission, given that $\mathcal{A}$ fails in the first transmission. Using (\ref{eq_SkH_2}), we have
\begin{equation}
s=\frac{1}{N-1} \sum_{i=0}^{N-1} \Big(k \times \operatorname{Pr}\left\{S_{k} \mid H\right\}\Big)=\frac{\operatorname{Pr}\left\{S_{1} \mid H\right\}}{N-1}.\nonumber
\end{equation}
It follows that $(N-1)(1-s)$ is the average number of UEs that do not compete for the same RB with UE $\mathcal{A}$ in $\mathcal{A}$'s first transmission. Each of these UEs will send a packet when $\mathcal{A}$ makes the first retransmission with probability $\alpha / 4$. Thus, we have
\begin{equation}
g_{1}=(N-1)(1-s) \frac{\alpha}{4}=(1-s) G_{F}.\label{eq_g1}
\end{equation}
As (\ref{eq_SkH_2}) indicates, $\mathcal{A}$ collides with at most 1 UE if it fails in the first transmission. If $\mathcal{A}$ collides with a UE and both of them fail, $\mathcal{A}$ will see this UE make a retransmission 4 TTIs later. In this case, this UE will contribute to $g_{2}$, which is the average number of UEs that fail in the competition with UE $\mathcal{A}$ in $\mathcal{A}$'s first transmission. Let $p_{1}$ be the probability that a UE fails in the competition with $\mathcal{A}$ when $\mathcal{A}$ sends a packet, given that $\mathcal{A}$ fails in the first transmission. According to Appendix A, we have
\begin{equation}
p_{1}=\left\{\begin{array}{cl}
\dfrac{1-\frac{2}{1+\mu} e^{-\frac{\mu}{\rho}}+\frac{1-\mu}{1+\mu} e^{-\frac{2 \mu}{(1-\mu) \rho}}}{1-\frac{1}{\mu+1} e^{-\frac{\mu}{\rho}}} & 0<\mu \leq 1 \\
\dfrac{1-\frac{2}{1+\mu} e^{-\frac{\mu}{\rho}}}{1-\frac{1}{\mu+1} e^{-\frac{\mu}{\rho}}} & \mu>1
\end{array}\right..\label{eq_p1}
\end{equation}
Combining (\ref{eq_p1}) with the fact that the failure of $\mathcal{A}$ in the first transmission is induced by the collision with at most one UE yields $g_{2}=0 \times \operatorname{Pr}\left\{S_{0} \mid H\right\}+1 \times p_{1} \operatorname{Pr}\left\{S_{1} \mid H\right\}$. As a result, the attempt rate seen by $\mathcal{A}$ in the first retransmission is equal to
\begin{align}
G_{R_{1}}&=g_{1}+g_{2}\nonumber\\&=\left\{1-\frac{\frac{G_{F}}{B}\left(1-\frac{e^{-\frac{\mu}{\rho}}}{\mu+1}\right)}{(N-1)\left[1-e^{-\frac{\mu}{\rho}}\left(1-\frac{\mu}{\mu+1} \frac{G_{F}}{B}\right)\right]}\right\} G_{F} \nonumber\\
&+\frac{p_{1} \frac{G_{F}}{B}\left(1-\frac{e^{-\frac{\mu}{\rho}}}{\mu+1}\right)}{1-e^{-\frac{\mu}{\rho}}\left(1-\frac{\mu}{\mu+1} \frac{G_{F}}{B}\right)}.\label{eq_GR1}
\end{align}
Following the similar argument, we can derive $G_{R_{i}}$ as follows
\begin{align}
G_{R_{i}}&=\left\{1-\frac{\frac{G_{R_{i-1}}}{B}\left(1-\frac{e^{-\frac{\mu}{\rho}}}{\mu+1}\right)}{(N-1)\left[1-e^{-\frac{\mu}{\rho}}\left(1-\frac{\mu}{\mu+1} \frac{G_{R_{i-1}}}{B}\right)\right]}\right\} G_{F}\nonumber \\
&+\frac{p_{1} \frac{G_{R_{i-1}}}{B}\left(1-\frac{e^{-\frac{\mu}{\rho}}}{\mu+1}\right)}{1-e^{-\frac{\mu}{\rho}}\left(1-\frac{\mu}{\mu+1} \frac{G_{R_{i-1}}}{B}\right)},\label{eq_GRi}
\end{align}
where $i=2,3, \cdots$.

To clearly demonstrate the relationship between $G_{F}$ and $G_{R_{i}}$, we consider the ideal wireless channel, where there is no noise \cite{no_noise}. In this case, the mean SNR $\rho=\bar{P} / \sigma^{2} \rightarrow \infty$ and $\mu / \rho \rightarrow 0$, and thus
\begin{equation}
G_{R_{i}}^{*}=\left(1-\frac{1}{N-1}\right) G_{F}^{*}+p_{1}^{*},\label{eq_GRi*}
\end{equation}
where  
\begin{equation}
p_{1}^{*}=\lim _{\rho \rightarrow \infty} p_{1}=\left\{\begin{array}{cc}
0 & \text { if } 0<\mu \leq 1 \\
1-1 / \mu & \mu>1
\end{array}\right..\nonumber
\end{equation}
In the case of ideal channel, Eq. (\ref{eq_GRi*}) shows that $G_{R_{1}}^{*}=G_{R_{2}}^{*}=\cdots=G_{R_{\infty}}^{*} \approx G_{F}^{*}+p_{1}^{*}$, where $p_{1}^{*}$ is actually the reduction of $g_2$ and is contributed by the UE who fails in the competition with $\mathcal{A}$ in the last transmission. This result is formally stated by the following theorem.

\begin{theorem}
If the number of RBs is sufficiently large such that the attempt rate on each RB is very small and the channel has no noise, all the attempt rates $G_{R_{1}}^{*}, G_{R_{2}}^{*}, \cdots$ seen by a UE in retransmissions are the same, and are approximately increased by $p_{1}^{*}$ when compared with the attempt rate $G_{F}^{*}$ seen by the UE in the first transmission.\label{Theorem}
\end{theorem}

Note that if the SINR threshold $\mu$ satisfies $0<\mu \leq 1$, the difference $p_{1}^{*}=0$, which means the UE in collision with $\mathcal{A}$ in the last transmission succeeds. This can be interpreted as follows. The ideal channel has no additive noise, i.e., $\sigma^{2}=0$. In this case, the SINRs of $\mathcal{A}$ and the UE in collision with $\mathcal{A}$ are reciprocal of each other, according to (\ref{eq_SINR}). It follows that given $\mathcal{A}$ fails, i.e., its SINR is less than 1, the SINR of the other UE must be larger than 1 and thus succeed in the competition.                

\begin{figure}[t]
\centerline{\includegraphics[width=0.6\linewidth]{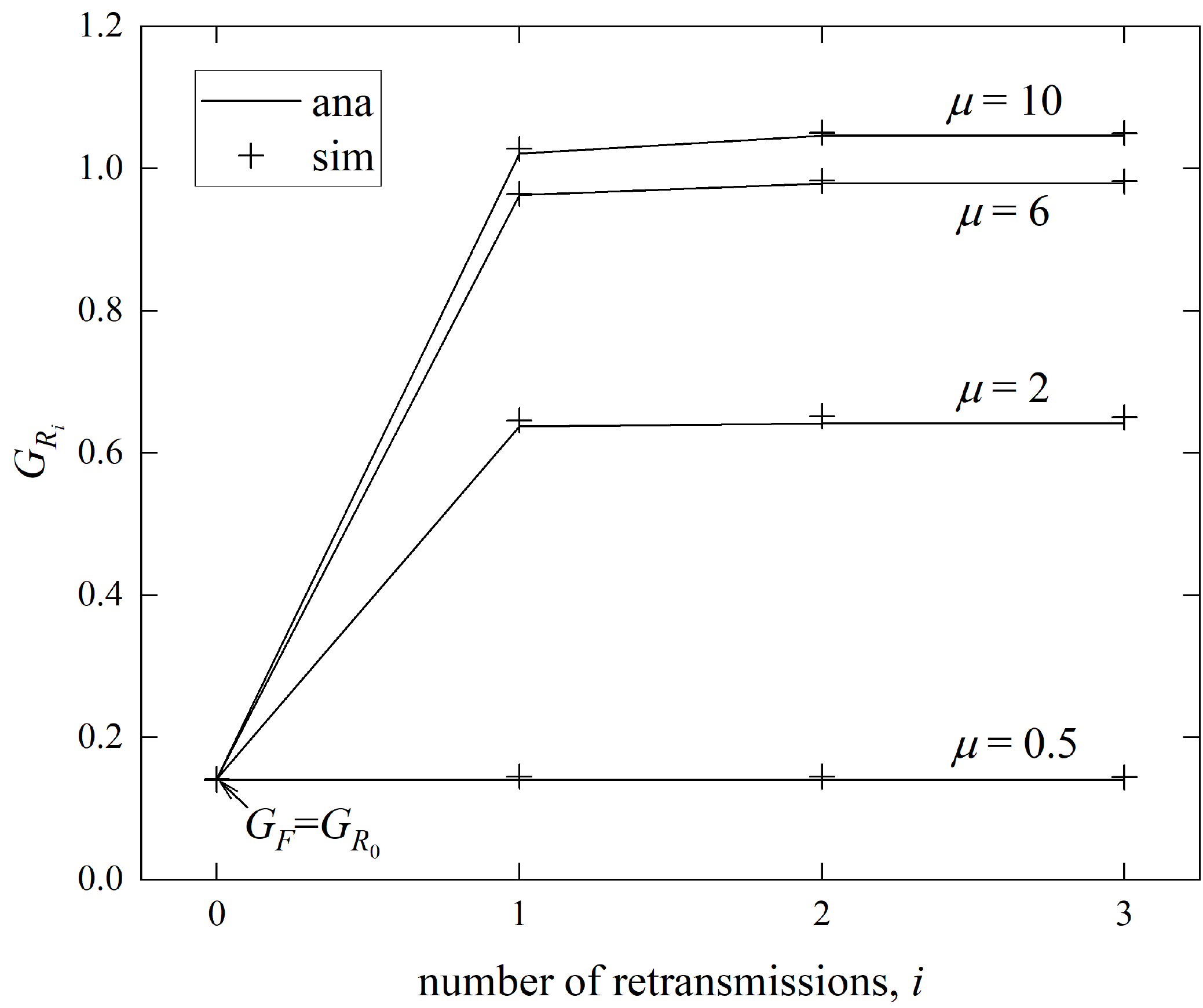}}
\caption{Effect of 1-pR on the attempt rates seen by a UE in different rounds of retransmission, where $\sigma^{2}=-112$dBm, $\bar{P}=-60$dBm, $N=100$, $B=48$ and $\lambda=10$packets/s.}\label{fig_attemptrate}
\end{figure}

As Eqs. (\ref{eq_GR1}) and (\ref{eq_GRi}) exhibit, the relationship between $G_F$ and $G_{R_i}$ in general cases is quite complicated. We thus use numerical results and simulations to demonstrate that Theorem 1 approximately holds. For a channel in general cases, it is necessary to keep the mean received SNR $\rho$ much larger than the SINR threshold $\mu$, i.e., $\mu / \rho \ll 1$, such that very high reliability of URLLC service can be guaranteed. Under this condition, Fig. \ref{fig_attemptrate} plots $G_{R_i}$ as a function of retransmission times $i$, where the UE population is $N=100$, the number of RBs is $B=48$, the average power of additive noise is $\sigma^{2}=-112$dBm, and the mean received power of packets is $\bar{P}=-60 \mathrm{dBm}$ \cite{transmit_power}. Note that $G_{R_0}$ in Fig. \ref{fig_attemptrate} stands for $G_F$, as we mark. From the result, we can find that the difference among the $G_{R_i}$s is much smaller than that between $G_{R_i}$ and $G_F$, except when $0<\mu \leq 1$.  

\section{Delay Performance Analysis} \label{sec_Delay}
This section studies the user-plane delay, denoted by $D$. Different from the previous works \cite{liuyan,similat_liuyan_reactive,similat_liuyan_K_rep,similat_liuyan_MIMO,Markov_chain_proactive}, we take into consideration the queueing process of packets. Following the idea of two-stage queueing model reported in \cite{non_empty_probability,two_stage,two_stage_2}, we model each UE as an M/G/1 queue, where the time from when the packet becomes an HOL packet to when the UE finishes processing an ACK from the BS is treated as the service time of a packet and denoted by $X$. Using the result of Theorem \ref{Theorem}, we first derive $X$ in section \ref{subsec_servicetime}, and then the distribution of $D$ in section \ref{subsec_US-Delay}.

\subsection{Service Time of an HOL Packet}\label{subsec_servicetime}
\begin{figure}[t]
\centerline{\includegraphics[width=0.6\linewidth]{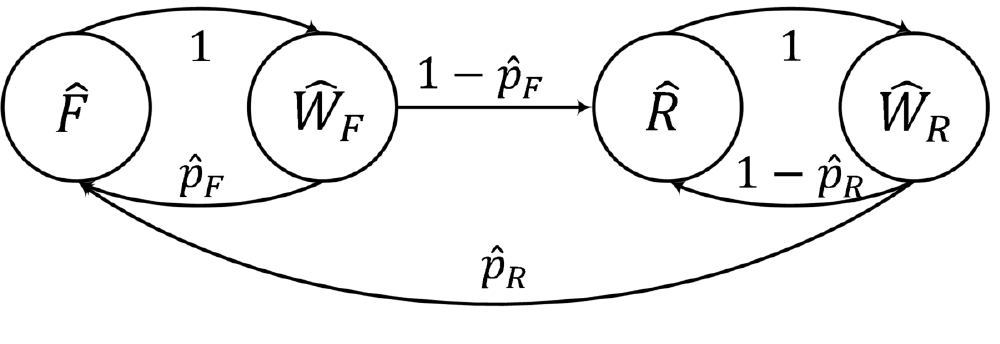}}
\caption{Simplified Markov chain of the competition process of an HOL packet.}\label{fig_S-MC}
\end{figure}

Section \ref{sec_1pR} demonstrates that all the $G_{R_i}$s are almost equal. This indicates that it is reasonable to adopt the following simplification:\\(*) \textit{All the attempt rates seen by the packet in different rounds of retransmission are all equal,}\\when we solve the service time, such that the derivation can be largely simplified. Accordingly, the Markov chain in Fig. \ref{fig_MC} reduces to a four-state Markov chain in Fig. \ref{fig_S-MC}, where there is a first-transmission state $\widehat{F}$, a waiting state after the first transmission $\widehat{W}_{F}$, a retransmission state $\widehat{R}$, and a waiting state after the retransmission $\widehat{W}_{R}$. In the simplified model, the attempt rates seen by the packet in the first transmission and the packet in retransmissions are denoted as $\widehat{G}_{F}$ and as $\widehat{G}_{R}$, respectively. In Fig. \ref{fig_S-MC}, $\widehat{p}_{F}$ and $\widehat{p}_{R}$ denote the successful probabilities of the first transmission and the retransmissions, respectively. 

According to the simplified Markov chain in Fig. \ref{fig_S-MC}, we can obtain the stationary probabilities of $\widehat{F}$, $\widehat{W}_{F}$, $\widehat{R}$, and $\widehat{W}_{R}$ as follows   
\[\pi_{\hat{F}}=\pi_{\widehat{W}_{F}}=\frac{\hat{p}_{R}}{2\left(1+\hat{p}_{R}-\hat{p}_{F}\right)},\]
and
\[\pi_{\hat{R}}=\pi_{\widehat{W}_{R}}=\frac{1-\hat{p}_{F}}{2\left(1+\hat{p}_{R}-\hat{p}_{F}\right)},\]
from which we further obtain the non-empty probability
\[\hat{\alpha}=\frac{2 \lambda}{\pi_{\hat{F}}}=\frac{4\left(1+\hat{p}_{R}-\hat{p}_{F}\right) \lambda}{\hat{p}_{R}}.\]
Accordingly, we have
\begin{equation}
\widehat{G}_{F}=(N-1) \hat{\alpha} / 4,\label{eq_GFhat}
\end{equation}
and
\begin{align}
\widehat{G}_{R}&=\left\{1-\frac{\frac{\widehat{G}_{F}}{B}\left(1-\frac{e^{-\frac{\mu}{\rho}}}{\mu+1}\right)}{(N-1)\left[1-e^{-\frac{\mu}{\rho}}\left(1-\frac{\mu}{\mu+1} \frac{\hat{G}_{F}}{B}\right)\right]}\right\} \hat{G}_{F}\nonumber\\&+\frac{p_{1} \frac{\widehat{G}_{F}}{B}\left(1-\frac{e^{-\frac{\mu}{\rho}}}{\mu+1}\right)}{1-e^{-\frac{\mu}{\rho}}\left(1-\frac{\mu}{\mu+1} \frac{\hat{G}_{F}}{B}\right)}.\label{eq_GRhat}
\end{align}
Following the argument that the packet can successfully transmitted if the received SINR is larger than the preset threshold $\mu$, we drive $\widehat{p}_{F}$ and $\widehat{p}_{R}$ according to (\ref{eq_GFhat}) and (\ref{eq_GRhat}) as follows 
\begin{equation}
\hat{p}_{F}=e^{-\frac{\mu}{\rho}}\left(1-\frac{\mu}{\mu+1} \frac{\widehat{G}_{F}}{B}\right),\label{eq_pFhat}
\end{equation}
and
\begin{equation}
\hat{p}_{R}=e^{-\frac{\mu}{\rho}}\left(1-\frac{\mu}{\mu+1} \frac{\widehat{G}_{R}}{B}\right).\label{eq_pRhat}
\end{equation}

In this case of ideal channel where $\mu / \rho \rightarrow 0$, (\ref{eq_GRhat}), (\ref{eq_pFhat}) and (\ref{eq_pRhat}) can be simplified as follows   
\begin{equation}
\hat{G}_{R}^{*}=\left(1-\frac{1}{N-1}\right) \hat{G}_{F}+p_{1}^{*}=\frac{\hat{\alpha}}{4}(N-2)+p_{1}^{*},\nonumber
\end{equation}
\begin{equation}
\hat{p}_{F}^{*}=1-\frac{\mu}{\mu+1} \frac{\hat{\alpha}}{4 B}(N-1),\nonumber
\end{equation}
and 
\begin{equation}
\hat{p}_{R}^{*}=\left(1-\frac{1}{N-1}\right) \hat{p}_{F}^{*}-\frac{\mu}{\mu+1} \frac{p_{1}^{*}}{B}.\nonumber
\end{equation}
These two equations indicate that $p_{R}^{*}$ is less than $p_{F}^{*}$ by a constant, which is the consequence of (\ref{eq_GRi*}).     

As Fig. \ref{fig_S-MC} shows, a UE may need several attempts for a successful transmission. The service time is 4 TTIs if the first transmission is successful with probability $p_{F}$, and $4+4i$ if the packet is sent successfully until the $i$-th retransmission with probability $\left(1-\hat{p}_{F}\right) \hat{p}_{R}\left(1-\hat{p}_{R}\right)^{i-1}$. It follows that the distribution of service time $X$ is given by \begin{equation}
\operatorname{Pr}\{X=x\}=\left\{\begin{array}{cc}
\hat{p}_{F} & x=4 \\
\left(1-\hat{p}_{F}\right)\left(1-\hat{p}_{R}\right)^{i-1} \hat{p}_{R} & x=4(i+1) \\
0 & \text { otherwise }
\end{array}\right.,\label{eq_X}
\end{equation}        
where $i=1,2,3 \cdots$.

\subsection{User-plane Delay Distribution}\label{subsec_US-Delay}
\begin{figure}[b]
\centerline{\includegraphics[width=1\linewidth]{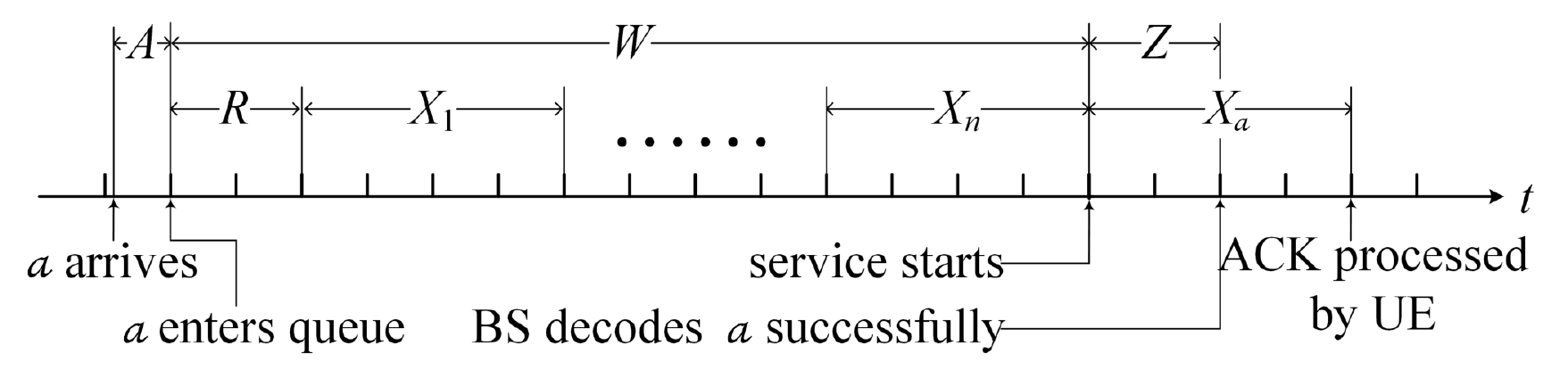}}
\caption{User-plane delay of packet $\mathcal{a}$ is equal to $D=A+W+Z$.}\label{fig_UP-Delay}
\end{figure}

As Fig. \ref{fig_UP-Delay} plots, the user-plane delay of packet $\mathcal{a}$ consists of the following parts. After arriving at the UE, $\mathcal{a}$ will experience an alignment delay, denoted by $A$, which is the time interval from the arrival to the start of the next TTI. After that, $\mathcal{a}$ enters the queue and experiences a waiting time, denoted by $W$, which includes the following two components. First, $\mathcal{a}$ has to wait a residual service time, denoted by $R$, if $\mathcal{a}$ sees a packet in service when it enters the queue. Second, $\mathcal{a}$ may need to wait the service completion of the packets waiting in the queue before $\mathcal{a}$. After $\mathcal{a}$ becomes an HOL packet, it will be successfully decoded by the BS after $X-2$ TTIs, called transmission time and denoted by $Z$. Thus, the user-plane delay is given by            
\begin{equation}
D=A+W+Z=A+W+X-2.\label{eq_D}
\end{equation}

The packet can be removed from the buffer of UE, only when the UE finishes processing an ACK from the BS. Let $V=W+X$ be the sojourn time that the packet spends in the queue. Let $d(t)$, $w(t)$ and $v(t)$ be the probability density functions (PDFs) of the user-plane delay, the waiting time and the sojourn time, the Laplace transforms of which are respectively denoted by $D(s)$, $W(s)$, and $V(s)$. According to (\ref{eq_D}), $D(s)=A(s)W(s)X(s)e^{2 s}=A(s)V(s)e^{2 s}$.

According to the P-K formula of M/G/1 queue \cite{p_k_formula}, $V(s)$ is given by 
\begin{equation}
V(s)=\frac{(1-\lambda \bar{X}) s}{s-\lambda+\lambda X(s)}X(s),\label{eq_V_Laplace}
\end{equation}
where $\bar{X} \triangleq E[X]=4\left(1+\hat{p}_{R}-\hat{p}_{F}\right) / \hat{p}_{R},$
and $X(s)$ is the Laplace transform of (\ref{eq_X})  
\begin{equation}
X(s)=\hat{p}_{F} e^{-4 s}+\frac{\left(1-\hat{p}_{F}\right) \hat{p}_{R} e^{-8 s}}{1-\left(1-\hat{p}_{R}\right) e^{-4 s}}.\label{eq_X_Laplace}
\end{equation}
Also, since the arrival process of packets at the UE is a Poisson process, $A$ is a uniformly distributed random variable in the range between 0 and 1 TTI, the Laplace transform of which is given by 
\begin{equation}
A(s)=\frac{1}{s}\left(1-e^{-s}\right).\label{eq_A_Laplace}
\end{equation}
Substituting (\ref{eq_V_Laplace}), (\ref{eq_X_Laplace}), and (\ref{eq_A_Laplace}) into $D(s)$  
\begin{equation}
D(s)=\frac{1}{s}\left(e^{2 s}-e^{s}\right) \frac{(1-\lambda \bar{X}) s}{s-\lambda+\lambda X(s)} X(s).\label{eq_D_Laplace}
\end{equation} 
We can obtain $d(t)$ through the inversion transform of (\ref{eq_D_Laplace}), and finally the delay distribution $D(t)$.  

\section{Numerical Discussions} \label{sec_discussion}
In the scenario where the group of UEs is relatively fixed in the coverage of a BS, we show that it is necessary to take into account the effect of 1-pR and queueing process of packets in the analytical model to provide an accurate performance prediction. We further show that this is especially important if we employ the analytical results to aid system design. We demonstrate this point through the comparison of analytical results and simulations, where we assume that the additive noise of wireless channel is $\sigma^{2}=-112$dBm, the average received power is $\bar{P}=-60$dBm \cite{transmit_power}, and the SINR threshold of the BS receiver is $\mu=4$dB.

\subsection{Effect of 1-pR on Model Accuracy}
As sections \ref{sec_1pR} and \ref{sec_Delay} discuss, the 1-pR will cause the attempt rate seen by the packet in retransmissions larger than that in the first transmission. It follows that the 1-pR will lower the successful probability of packet retransmission, which will enlarge the distribution tail of user-plane delay. This implies that whether to consider the effect of 1-pR in the analysis would have a significant impact on the accuracy of the analytical results.

To better verify this point, we first derive the distribution of user-plane delay without considering the effect of 1-pR for comparison. Ignoring the effect of 1-pR is equivalent to set $\hat{p}_{R}=\hat{p}_{F}=\tilde{p}$ in Fig. \ref{fig_MC}. Following the similar derivation process in section \ref{sec_Delay}, we have
\begin{equation}
\tilde{p}=\frac{1}{2} e^{-\frac{\mu}{\rho}}\left(1+\sqrt{1-\frac{4 e^{\frac{\mu}{\rho}}(N-1) \mu \lambda}{B(\mu+1)}}\right),\nonumber
\end{equation}
and 
\begin{equation}
\widetilde{D}(s)=\frac{1}{s}\left(e^{2 s}-e^{s}\right) \frac{(1-4 \lambda / \tilde{p}) s}{s-\lambda+\lambda \tilde{X}(s)} \widetilde{X}(s),\label{eq_D_tilde_Laplace}
\end{equation}
where
\begin{equation}
\tilde{X}(s)=\frac{\tilde{p} e^{-8 s}}{1-(1-\tilde{p}) e^{-4 s}},\nonumber
\end{equation}
is the Laplace transform of the PDF of the service time when the effect of 1-pR is ignored.

\begin{figure}[t]
\centering
\subfigure[$\lambda=5$ packets/s]{
\includegraphics[width=0.6\linewidth]{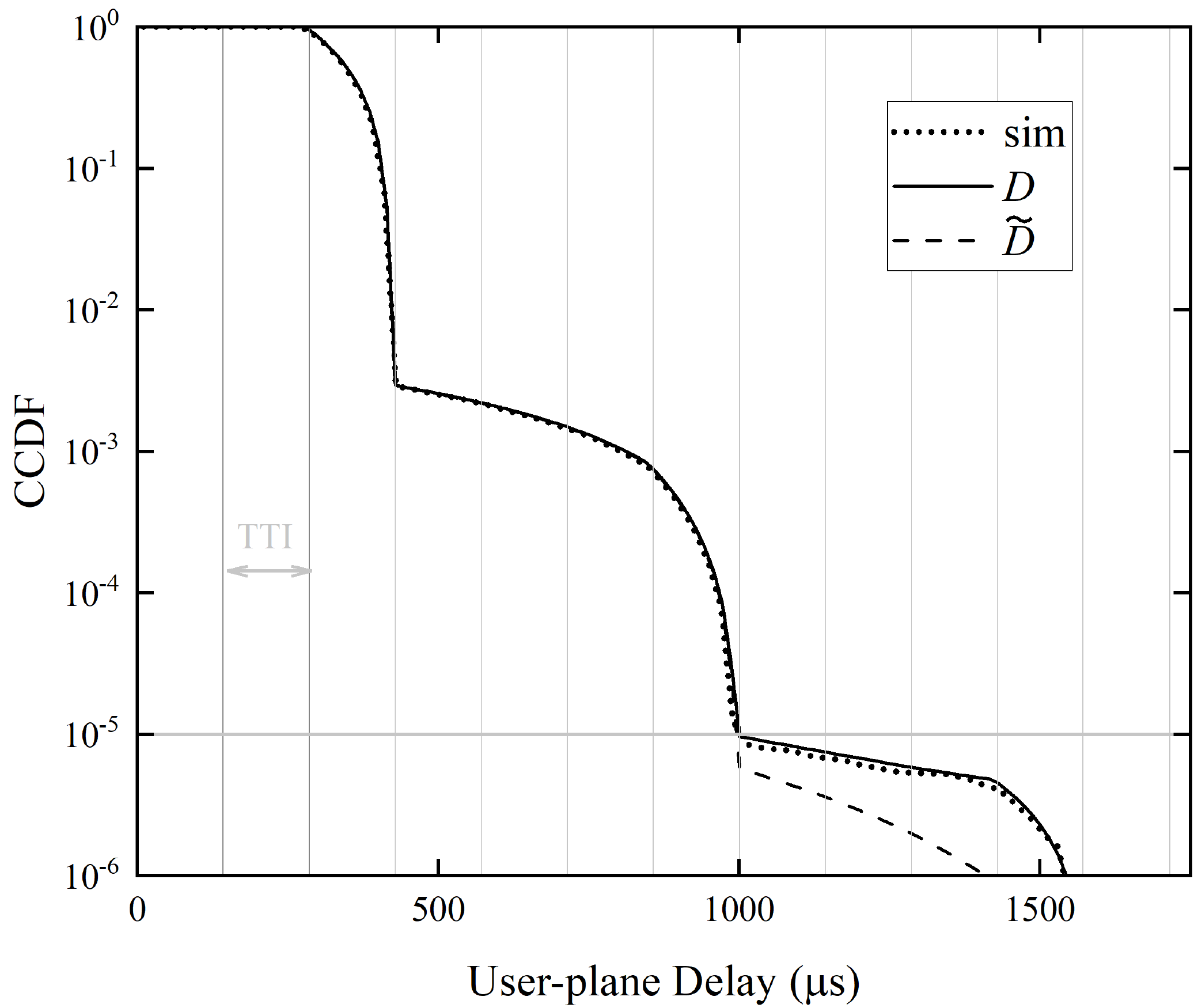}
\label{fig_Eff_1pR_5}
}
\quad
\subfigure[$\lambda=10$ packets/s]{
\includegraphics[width=0.6\linewidth]{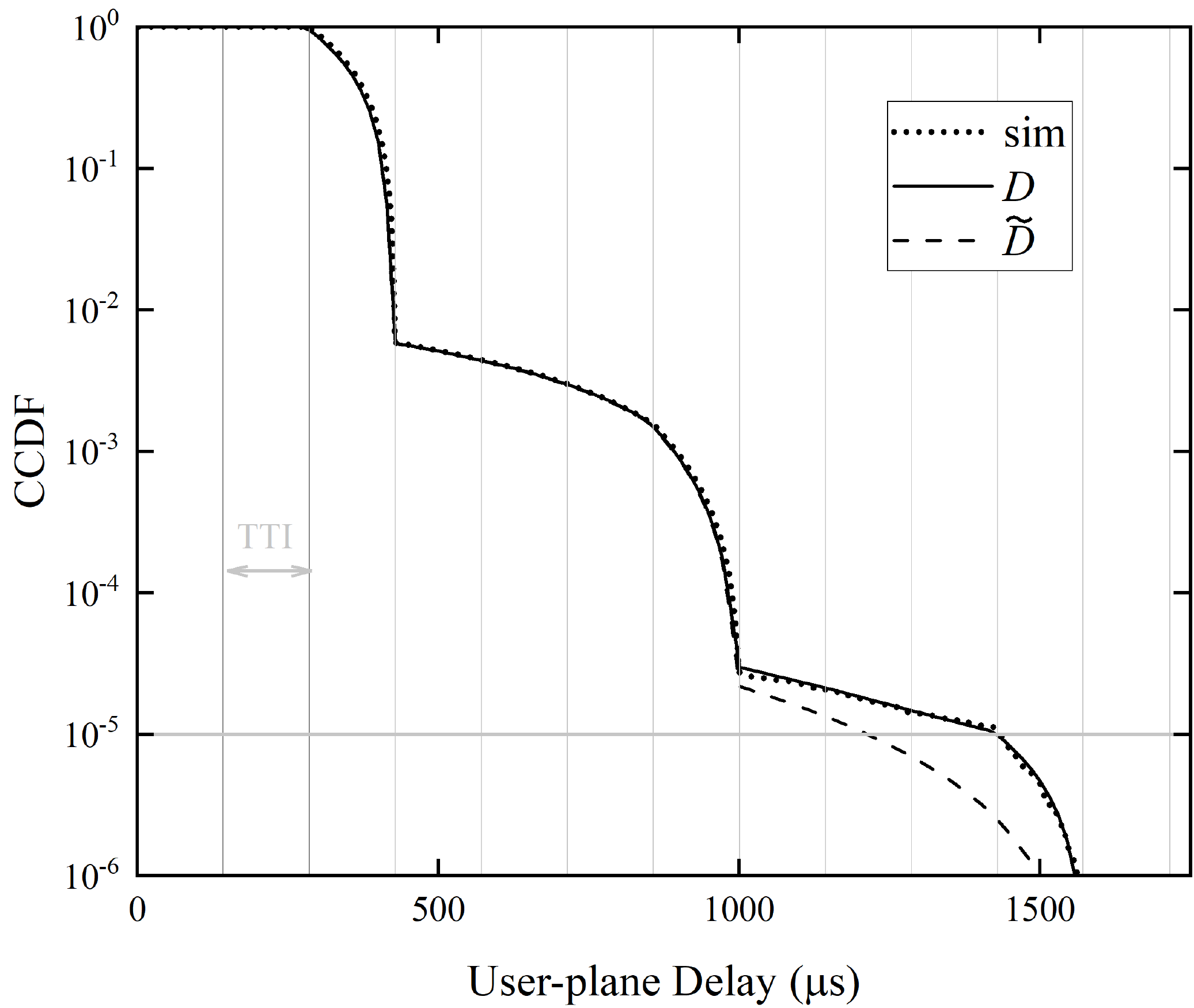}
\label{fig_Eff_1pR_10}
}
\caption{Effect of the 1-pR on delay performance, where $N=40$, $B=48$, $\sigma^{2}=-112$dBm, $\bar{P}=-60$dBm and $\mu=4$dB.}\label{fig_Eff_1pR}
\end{figure}

Using (\ref{eq_D_Laplace}) and (\ref{eq_D_tilde_Laplace}), Fig. \ref{fig_Eff_1pR} compares the complementary cumulative distribution functions (CCDFs) of user-plane delay when the effect of 1-pR is or is not considered in the analysis. The input traffic rates to a UE in Fig. \ref{fig_Eff_1pR_5} and \ref{fig_Eff_1pR_10} are 5 and 10 packets/s. As Fig. \ref{fig_Eff_1pR} plots, the curves are roughly stair-like, and the $i$-th step describes the delay distribution of packets that are transmitted $i$ times before success. Since the 1-pR is considered in the derivation of (\ref{eq_D_Laplace}), our analytical result $D$ agrees with the simulation result very well. As a comparison, $\widetilde{D}$ in (\ref{eq_D_tilde_Laplace}) that is obtained by assuming $\hat{p}_{R}=\hat{p}_{F}$ has a remarkable error on the second step. This is attributed to the reason that the assumption $\hat{p}_{R}=\hat{p}_{F}$ overestimates the successful probability of retransmission, which leads to the curve of $\widetilde{D}$ is lower than the simulation curve in the second step. This clearly indicates that taking into consideration the effect of 1-pR in delay analysis is necessary.  

\subsection{Effect of Queueing Process on Model Accuracy}
Though the arrive rate of URLLC traffic is very low, it is still necessary to consider the effect of queueing process on the user-plane delay. We demonstrate this point by comparing the delay performance with and without considering the queueing process. Let $D_{n q}$ be the user-plane delay when the arriving packet is assumed to see the UE with empty queue. The user-plane delay in this case consists of the alignment delay $A$ and the transmission delay $Z$, i.e., $D_{n q}=A+Z=A+X-2$. Let $d_{n q}(t)$ be the PDF of $D_{n q}$. Using (\ref{eq_X_Laplace}) and (\ref{eq_A_Laplace}), we have        
\begin{equation}
d_{n q}(t)=\left\{\begin{array}{cc}
\hat{p}_{F} & 2 \leq t<3 \\
\left(1-\hat{p}_{F}\right)\left(1-\hat{p}_{R}\right)^{i-1} \hat{p}_{R} & 2+4 i \leq t<3+4 i \\
0 & \text { otherwise }
\end{array}\right.\label{eq_Dnq}
\end{equation}
where $i=1,2,3 \cdots$. From (\ref{eq_Dnq}), we can obtain $D_{n q}(t)$.

\begin{figure}[t]
\centering
\subfigure[$\lambda=5$ packets/s]{
\includegraphics[width=0.6\linewidth]{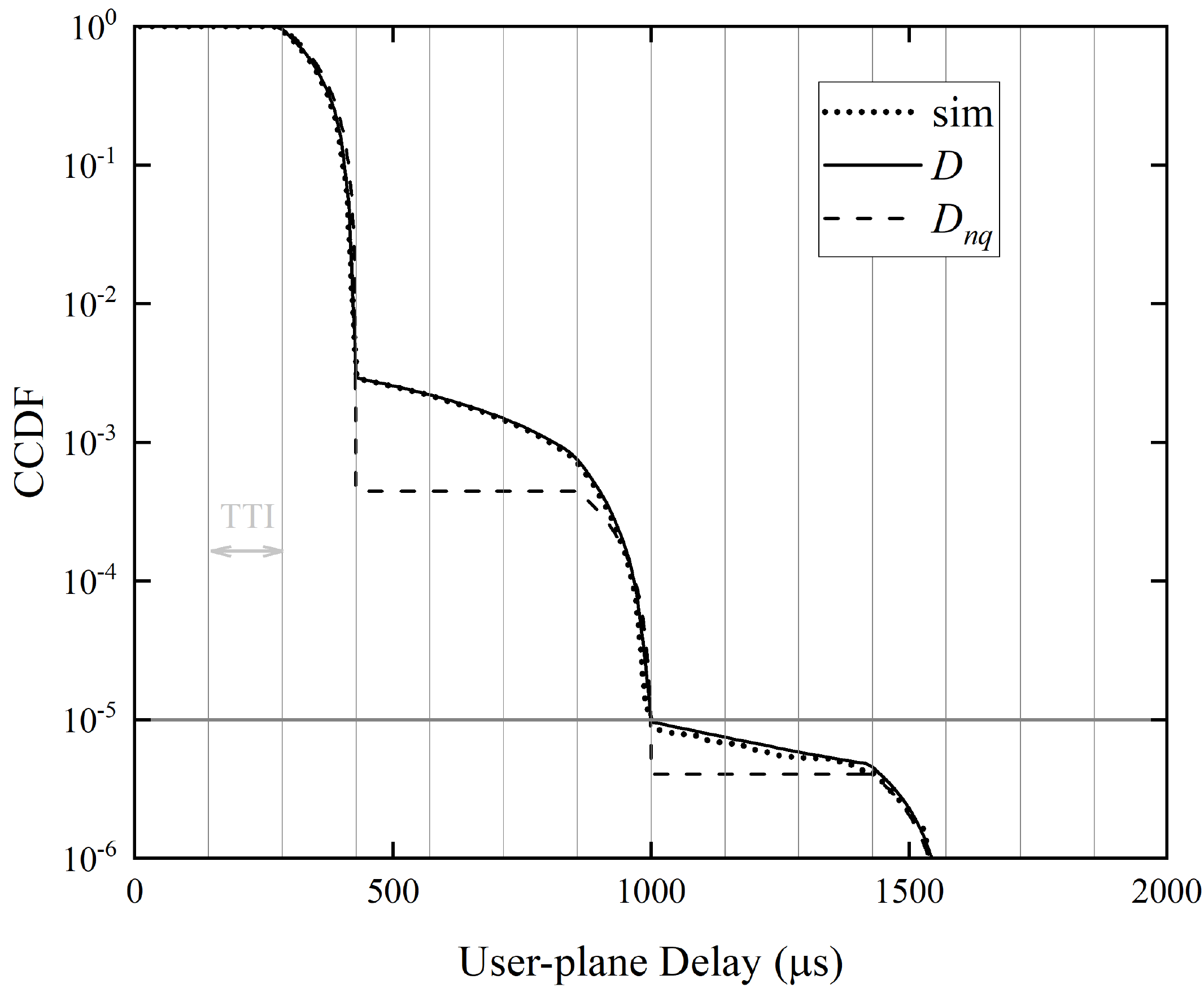}
\label{fig_queue_5}
}
\quad
\subfigure[$\lambda=10$ packets/s]{
\includegraphics[width=0.6\linewidth]{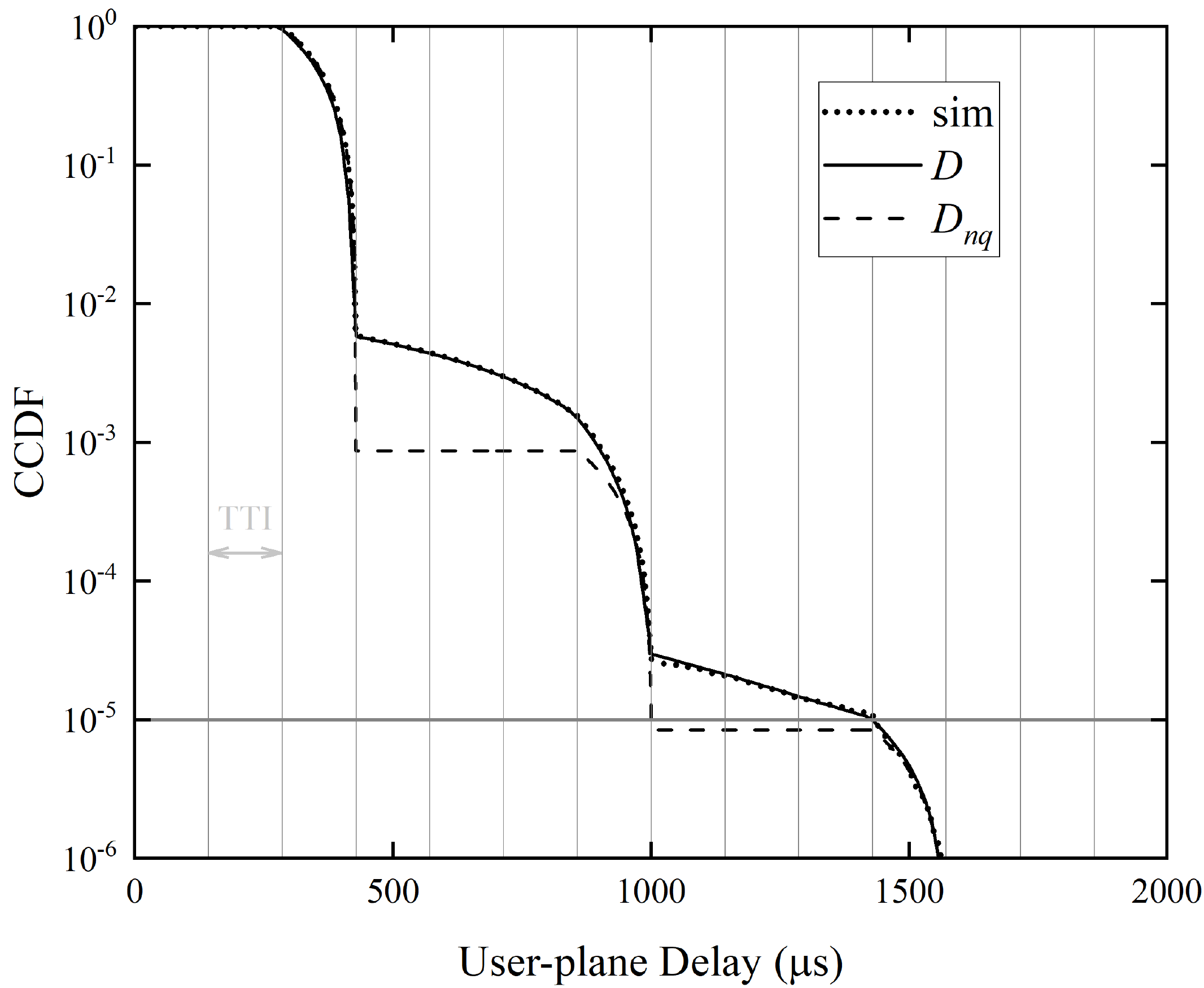}
\label{fig_queue_10}
}
\caption{Effect of the queueing process on delay performance, where $N=40$, $B=48$, $\sigma^{2}=-112$dBm, $\bar{P}=-60$dBm and $\mu=4$dB.}
\label{fig_queue}
\end{figure}

We compare $D_{n q}$, $D$, and simulation in Fig. \ref{fig_queue}, where the input traffic rates to a UE are 5 and 10 packets/s. A packet will have to wait several TTIs in the buffer before transmission if it sees a nonempty queue when it arrives at the UE. Therefore, the stair-like curves of the simulation result and $D$ are sloped in each step, as Fig. \ref{fig_queue} plots. However, if we assume that the arriving packet always sees an empty UE, the user-plane delay will not take some specific values. For example, the user-plane delay in this case will not take values like 4 and 5, as (\ref{eq_Dnq}) shows. Accordingly, in Fig. \ref{fig_queue}, there includes a horizontal line segment in each step of the curve of $D_{n q}$, e.g., the horizontal line segment from the 3rd TTI to the 6th TTI in the first step. This makes the curve of $D_{n q}$ remarkably lower than that of the simulation and $D$ at each step. 

\begin{figure}[t]
\centerline{\includegraphics[width=0.6\linewidth]{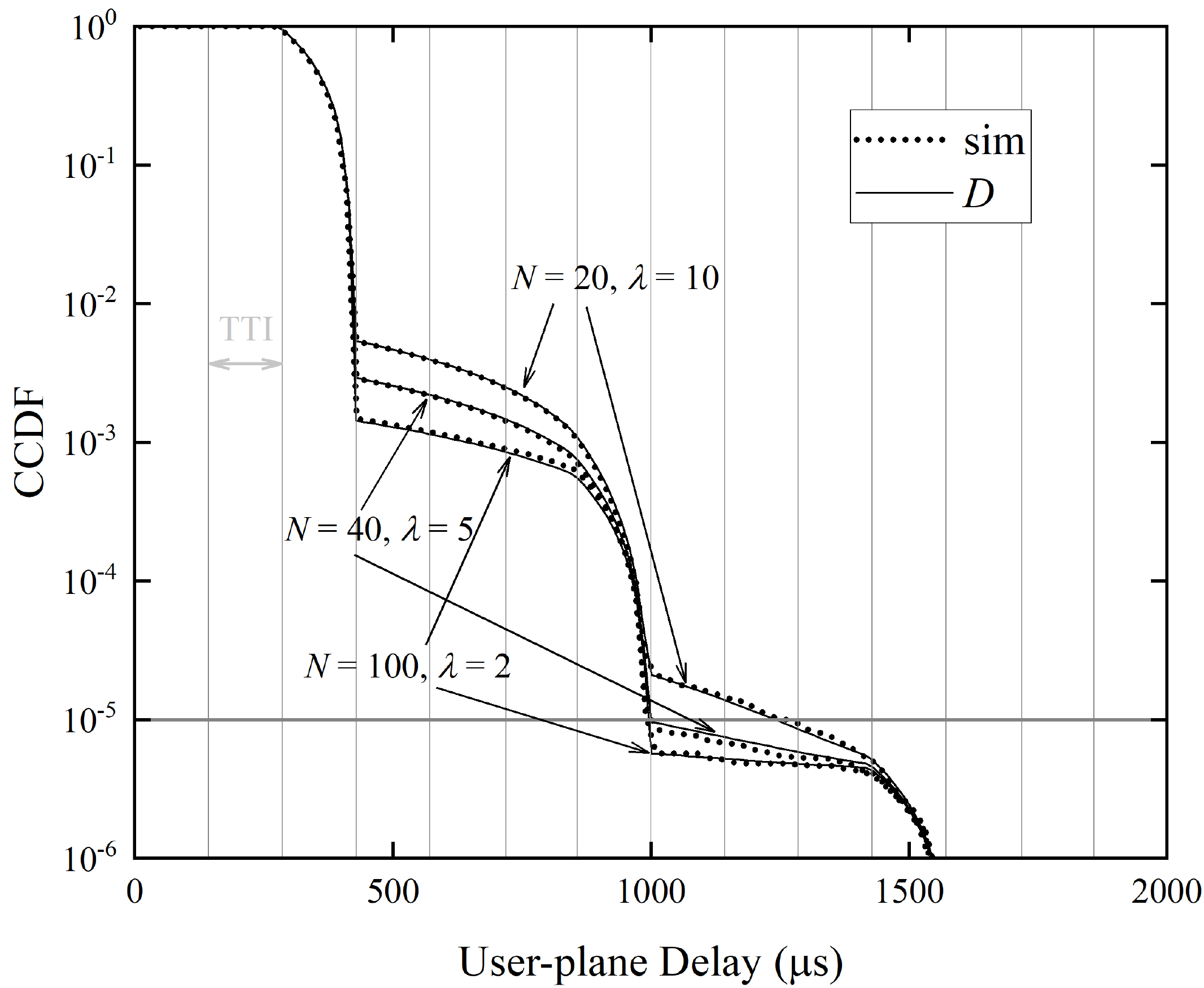}}
\caption{CCDF of user-plane delay under different $\lambda$s, where $\hat{\lambda}=200$ packets/s, $B=48$ and $\mu=4$dB.}
\label{fig_diff_lambda}
\end{figure}

We further observe the impact of queueing process on user-plane delay under different input traffic rates to a UE, while fixing the aggregate input traffic rate $\hat{\lambda}=200$ packets/s. It is very interesting to see from Fig. \ref{fig_diff_lambda} that the curve of small $N$ and relatively large $\lambda$ is higher than that of large $N$ and relatively small $\lambda$ in each step. This clearly indicates that the input traffic rate has a greater impact on the delay performance than the UE population, because an arriving packet in the case with relatively large $\lambda$ and small $N$ will have more chance to see a non-empty UE and hence suffer a larger queueing delay. In particular, the access network in the case with $\lambda=10$ packets/s and $N=20$ already cannot meet the reliability requirement that the user-plane delay of $>99.999\%$ packets should be less than 1ms. This suggests that, due to the impact of queueing process, we should consider the traffic rate to each UE $\lambda$, not just the aggregate rate $\hat{\lambda}$, in the system design.

\subsection{Applications in System Design}
Compared to the simulation, the analytical result can obtain the outage probability very fast, such that it can be used to assist the design of networks. In practice, the BS is able to know if there is a UE that newly joins or departs from its coverage, via Dual Active Protocol Stack \cite{cell_handover}. Hence, a possible application of analytical results is to assign a proper number of RBs, denoted by $B^{*}$, to the URLLC traffic, according to the channel state, the number of UEs, and the input traffic rate to a UE . However, if the analytical model is not accurate enough, it will exert a negative influence on system configuration, as we demonstrate below.   

The outage probability, denoted by $P_{o}$, is defined as the probability that the user-plane delay of a packet is larger than a preset threshold. According to the 3GPP standard \cite{3gpp_standard}, $P_{o}$, must be less than $10^{-5},$ given that the delay threshold is 1ms. Clearly, the outage probability can be easily calculated from the PDF of user-plane delay. For example, we can find the outage probability using $d(t)$:     
\begin{equation}
P_{o}=1-\int_{0}^{7} d(t) dt.\nonumber
\end{equation}

Assuming that $\sigma^{2}$, $\bar{P}$, $N$, and $\lambda$ are known, we find the proper number of RBs $B^{*}$ via a simple method as follows. Initially, select a $B$ and calculate $P_{o}$. If $P_{o}>10^{-5}$, increase $B$ by a step size $b<B$ and recalculate $P_{o}$. Repeat the process until $P_{o}<10^{-5}$. We plot $P_{o}$ as a function of $B$, and find the intersection of the curve with the line $P_{o}=10^{-5}$. The rounding up of the abscissa of the intersection point is the $B^{*}$ to be found. If the initial $P_{o}$ is small than $10^{-5}$, we can find $B^{*}$ in the similar way. Table \ref{tab_B*} compares the $B^{*}$ predicted by simulation, $D$, $\tilde{D}$, and $D_{n q}$, where $\lambda=5$ packets/s and $N$ changes from 40 to 100. It can be seen that $D$ can provide an exact prediction, while $\widetilde{D}$ and $D_{n q}$ lead to a remarkably underestimated $B^{*}$. Clearly, the underestimation of $B^{*}$ will lead to a bad consequence that the reliability cannot be guaranteed. For example, when $N=60$, the $B^{*}$s predicted by $\widetilde{D}$ and $D_{n q}$ are 29 and 39, respectively. If we employ these predicted $B^{*}$s, our simulation shows that the outage probabilities would be $2.4 \times 10^{-5}$ and $1.7 \times 10^{-5}$, respectively, which are higher than $10^{-5}$.                                                           
\begin{table}[t]
\caption{$\sigma^{2}=-112$\textup{dBm}, $\bar{P}=-60$\textup{dBm}, and $\lambda=5$\textup{packets/s}}
\begin{center}
\begin{tabular}{|c|c|c|c|c|}
\hline
$N$&$B^{*}$ by sim&$B^{*}$ by $D$&$B^{*}$ by $\widetilde{D}$&$B^{*}$ by $D_{n q}$   \\
\hline\hline
40&47&47&20&32    \\
\hline
60&60&60&29&39    \\
\hline
80&73&73&39&45  \\
\hline
100&84&84&49&52  \\
\hline
\end{tabular}
\end{center}\label{tab_B*}
\end{table}

Note that we also consider the case where $\lambda=10$ packets/s. We fail to find a $B^{*}$ such that the network can support, no matter what value $N$ takes and whether we use the simulation or the analytical model. This is because a considerable portion of packets will experience the queueing process. In this case, some additional mechanism should be taken to eliminate the queue delay, such as 4 stop-and-wait (SAW) channel discussed in \cite{0.143_figure}. 

\subsection{Case Study: A Dynamic Scenario}
Our model is developed under the assumption that the group of UEs in the coverage of a BS is fixed. However, the UE in reality may move from one cell to another. We thus verify the effectiveness of our model in a dynamic scenario, where the UE group changes slowly.

To do that, we make the following assumptions. We assume the radius of a cell is 250m \cite{BS_radius}, and the moving speed of a UE is 30km/h, meaning that a UE will leave a cell with probability $p_{l}=2.3 \times 10^{-6}$ in each TTI. Let $N_{t}$ be the number of UEs in TTI $t$ and $\bar{N}$ be the average number of UEs in a cell. Suppose that the average number of UEs moving into a cell is $\left(2 \bar{N}-N_{t}\right) p_{l}$, such that $N_{t}$ fluctuates around $\bar{N}$. At last, we assume that the channel condition keeps unchanged.

To utilize the analytical results in such a dynamic scenario, we calculate the $B^{*}$s for all possible $N$s beforehand and store them in a table similar to Table \ref{tab_B*}. Once the BS detects the variation of $N_{t}$, it adjusts $B^{*}$ according to the table, such that spectrum utilization can be enhanced.         

We perform simulation according to the above description. Table \ref{tab_p_o} gives the outage probability when the BS adjusts the number of RBs according to $N_{t}$, using $D$, $\widetilde{D}$ and $D_{n q}$, and confirms that the adjustment based on $D$ can ensure an outage probability of less than $10^{-5}$ in the dynamic situation.           

\begin{table}[t]
\caption{$\sigma^{2}=-112$\textup{dBm}, $\bar{P}=-60$\textup{dBm}, and $\lambda=5$\textup{packets/s}}
\begin{center}
\begin{tabular}{|c|c|c|}
\hline
$p_o$ of $D$&$p_o$ of $\widetilde{D}$&$p_o$ of $D_{n q}$   \\
\hline\hline
$9.24 \times 10^{-6}$&$3 \times 10^{-5}$&$1.65 \times 10^{-5}$     \\
\hline
\end{tabular}
\end{center}\label{tab_p_o}
\end{table}
\section{Conclusion}\label{sec_conclusion}
This paper studies the impact of 1-pR on the competition process of packets when the UE group under the coverage of a BS is relatively fixed. We derive the distribution of user-plane delay, taking into account the queueing process. We verify our analytical result via simulation. According to our analysis, we have the following findings:
\begin{enumerate}[label=\alph*)]
\item 	The 1-pR increases the attempt rate seen by retransmitted packets, and thus reduces the successful probability of retransmission;
\item   The user-plane delay will be underestimated if the analytical model does not consider the effect of 1-pR;
\item 	Though only a few packets suffer queueing process, it is already enough to affect the delay performance.
\end{enumerate}
We further apply our analytical result to system configuration, and show that it can ensure an outage probability of less than $10^{-5}$. In the future, we will extend our model to analyze the Reactive with multiple SAW channels and the K-repetition.

\begin{appendices}
\setcounter{equation}{0}
\renewcommand{\theequation}{A-\arabic{equation}}
\section{Derivation of $p_{1}$}
Recall that $p_{1}$ is the probability that a UE, denoted by $\mathcal{B}$, fails in the competition with a specific UE $\mathcal{A}$ when $\mathcal{A}$ transmits the HOL packet for the first time, given that $\mathcal{A}$ fails in the first transmission, $S_1$ is the event that $\mathcal{B}$ competes with $\mathcal{A}$ for the same RB when $\mathcal{A}$ sends its HOL packet for the first time, and $H$ is the event that $\mathcal{A}$ fails in the first transmission of HOL packet. Let $E$ be the event that a UE fails in the competition with $\mathcal{A}$. $p_{1}$ can be represented as follows:
\begin{equation}
p_{1}=\operatorname{Pr}\left\{E \mid H, S_{1}\right\}=\frac{\operatorname{Pr}\left\{E, H \mid S_{1}\right\}}{\operatorname{Pr}\left\{H \mid S_{1}\right\}},\label{eq_p1_1}
\end{equation}
$\operatorname{Pr}\left\{E, H \mid S_{1}\right\}$ is the probability that $\mathcal{B}$ and $\mathcal{A}$ compete for the same RB and both of them fail in packet transmission. Let $\left|h_{\mathcal{A}}\right|^{2}$ and $\left|h_{\mathcal{B}}\right|^{2}$ be the channel power gains of $\mathcal{A}$ and $\mathcal{B}$, respectively. According to \cite{liuyan}, the channel power gain is an exponential distributed random variable with parameter 1. Assume that $\left|h_{\mathcal{A}}\right|^{2}$ and $\left|h_{\mathcal{B}}\right|^{2}$ are independent. The joint PDF of $\left|h_{\mathcal{A}}\right|^{2}$ and $\left|h_{\mathcal{B}}\right|^{2}$ is given by $f\left(\gamma_{1}, \gamma_{2}\right)=e^{-\gamma_{1}-\gamma_{2}}$, where $\gamma_{1}>0$ and $\gamma_{2}>0$. It follows that          
\begin{equation}
\operatorname{Pr}\left\{E, H \mid S_{1}\right\}=\iint_{U} f\left(\gamma_{1}, \gamma_{2}\right) d\gamma_{1}d\gamma_{2},\label{eq_EHS1}
\end{equation}
where the domain of integration $U$ is defined by following inequalities:  
\begin{equation}
\frac{\bar{P} \gamma_{1}}{\sigma^{2}+\bar{P} \gamma_{2}} \leq \mu,\nonumber
\end{equation}
and
\begin{equation}
\frac{\bar{P} \gamma_{2}}{\sigma^{2}+\bar{P} \gamma_{1}} \leq \mu,\nonumber
\end{equation}
which means both $\mathcal{A}$ and $\mathcal{B}$ fail because of insufficient received SINR. 
Substituting (\ref{eq_HSk}) with $k=1$ and (\ref{eq_EHS1}) into (\ref{eq_p1_1}), we finally obtain (\ref{eq_p1}) in section \ref{sec_1pR}.

\end{appendices}

\bibliographystyle{IEEEtran}
\bibliography{IEEEabrv,ref}

\end{document}